\documentclass[aps,prd,reprint,groupedaddress]{revtex4}
\usepackage{amsmath}
\usepackage{graphicx}
\usepackage{euscript}
\usepackage{braket}
\usepackage{euscript}

\begin{document}
	\title{Small $x$ phenomenology on gluon evolution through BFKL equation in light of a constraint in multi-Regge kinematics }
	
	
	\author{Pragyan Phukan}
	\email[]{pragyanp@tezu.ernet.in}
	\affiliation{HEP laboratory, Department of Physics, Tezpur University, India}
	\author{Madhurjya Lalung}
	\email[]{mlalung@tezu.ernet.in}
	\affiliation{HEP laboratory, Department of Physics, Tezpur University, India}
	
	\author{Jayanta Kumar Sarma}
	\email[]{jks@tezu.ernet.in}
	\affiliation{HEP laboratory, Department of Physics, Tezpur University, India}
	
	
	\begin{abstract}
		We investigate the impact of so called kinematic constraint on gluon evolution at small $x$. Implanting the constraint on the real emission term of gluon ladder diagram, we obtain an integro-differential form of BFKL equation. Later we solve the equation analytically using the method of characteristics. We sketch the Bjorken x and transverse momentum $k_t^2$ dependence of our solution of unintegrated gluon distributions $f(x,k_t^2)$ in  kinematic constraint supplemented BFKL equation and contrasted the same with original BFKL equation. Then we  extract the collinear gluon density $xg(x,Q^2)$ from unintegrated gluon distributions $f(x,k_t^2)$ and compared our theoretical prediction with that of global data fits viz. NNPDF3.1sx and CT14. Finally we assess the sensitivity of $f(x,k_t^2)$ towards BFKL intercept $\lambda$ for three canonical choices of $\lambda$ viz. o.4, 0.5 and 0.6.
	\end{abstract}
	
	\maketitle
	
	\section{Introduction}\label{intro}
\begin{figure*}[t]
		\centering 
		\includegraphics[width=.46\textwidth]{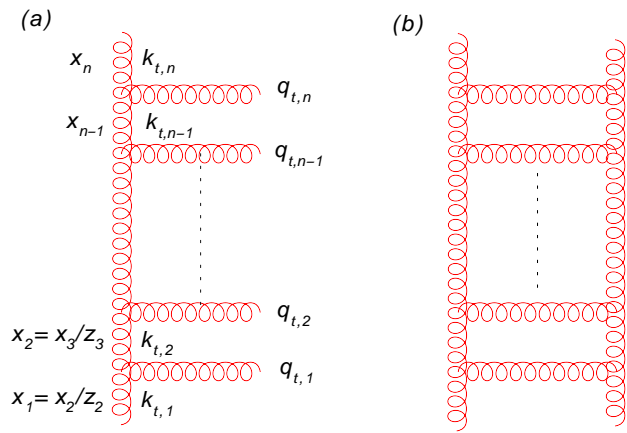}
		
		\caption{\label{f1}(a) Chain of sequential gluon emission which forms the basis of BFKL equation. On squaring the amplitude of (a) the ladder (b) is generated which when summed gives the BFKL kernel.}
	\end{figure*}

	The parton distribution functions (PDFs) serves as a very important tool for the calculation of inclusive cross sections in hadronic collision processes. In perturbative QCD, particularly for moderate values of Bjorken $x$ and large interaction scale $Q^2$, DIS observables are determined by the mass factorization theorem in which the collinear logarithmic singularities arising from gluon splitting are characterized by collinear parton distribution functions. These collinear PDFs are determined by the DGLAP evolution equation \cite{63,64,65} which resums the leading $\alpha_s\ln Q^2$ contributions associated with a space like chain of n gluons with strongly ordered successive gluon transverse momentum along the chain. On the other hand, at sufficiently high energies ($\sqrt{S}$) or small $x$, the leading logarithmic contribution of $\alpha_s \ln (1/x)$ cannot be neglected along the gluon chain. In this regime, the dominant parton is the gluon and the key ingredient to calculate the scattering observables is the unintegrated gluon distributions (ugd's) entering through $k_t$ factorization formula. The ugd's are characterised by BFKL evolution \cite{66,67} which sums the leading  $\alpha_s \ln (1/x)$ contributions, usually written in the form
	\begin{equation}
	\label{1}
	\begin{split}
	-x\frac{\partial f \left(x, k_t^2\right)}{\partial x}=\frac{\alpha_s N_c k_t^2}{\pi}\int_{k_{0}^{'2}}^{\infty}\frac{dk_t^{'2}}{k_t^{'2}}\left[\frac{f (x, k_t^{'2})-f \left(x, k_t^2\right)}{|k_t^{'2}-k_t^2|} + \frac{f \left(x, k_t^2\right)}{\sqrt{k_t^4+4k_t^{'4}}}\right],
	\end{split}
	\end{equation}
	where $f \left(x, k_t^2\right)$ denotes the gluon density unintegrated over the gluon transverse momentum $k_t$ and $k_{0 }^{'2}$ is the infrared cutoff of the evolution. To gain insight of \eqref{1} we refer to Fig.~\ref{f1}(a) which shows a chain of sequential gluon emission. The BFKL kernel in \eqref{1} corresponds to the sum of gluon ladder diagram Fig.~\ref{f1}(b) which is formed by squaring the amplitude of Fig.~\ref{f1}(a). The real emission is characterized by the first term in the kernel while the second term corresponds to the diagrams with virtual corrections. The apparent singularity observed at $k_t^{'2}=k_t^2$ cancels between real and virtual contributions to ensure that there is no overall singularity in \eqref{1}.
	\par The real emission term in the BFKL kernel interpretes the $k'\rightarrow k+q$ splitting inside the hadron as shown in Fig.~\ref{f1}(a). In BFKL multi-regge kinematics the longitudinal component of the gluon momentum is strongly ordered $x_1\ll x_2\text{.....}\ll x_{n-1}\ll x_n$ whereas there is no ordering in the transverse component $k_{1 t}\sim k_{2 t}\sim \text{...}.\sim k_{\text{n-1} t}\sim k_{\text{n}t}$ and the virtuality of the gluon comes dominantly from the transverse component of the momentum i.e.
	\begin{equation}
	\label{2.3}
	k^2=2 k^+ k^--k_t^2\approx k_t^2.
	\end{equation}  
	As a consequence of these BFKL kinematics, so-called kinematic constraint or consistency constraint comes into the picture and it is implemented in different forms: 
	\begin{align}
	\label{1.1}
	q_t^2&<\frac{k_t^2}{z} \text{     } \text{ LDC \cite{6,74}},\\
	\label{1.2}
	q_t^2&<\frac{(1-z)k_t^2}{z}\text{  \cite{3}},\\
	\label{1.3}
	k_t^{'2}&<\frac{k_t^2}{z}\text{      }\text{  BFKL \cite{3}}.
	\end{align}
	The relation \eqref{1.3}  can be considered as a special case of \eqref{1.1} owing to the fact that for a fixed value of $k_t$, a high $q_t^2$ would imply an equally high $k_t^{'2}$ \cite{3}. Although there exist other constraints consequence of energy momentum conservation, this bound is considerably tighter than the later\cite{3}.
	
	\par Our primary goal of this work is to investigate the effect of this kinematic constraint on the small $x$ regime of gluon evolution. Accordingly we organise the content of the paper as follows. In Sect.~\ref{theory} we suggest an analytical solution to the kinematic constraint supplemented  BFKL equation. We adopt method of characteristics to solve the PDE. Then in Sect.~\ref{result} we present the numerical analysis and portray our results for gluon evolution. First we sketch both the small $x$ and transverse momentum $k_t^2$  dependence of unintegrated gluon distributions $f(x,k_t^2)$. Then we  extract collinear gluon density $xg(x,Q^2)$ from ugd's and study $x$ and $Q^2$ dependence of $xg(x,Q^2)$. Our theoretical predictions for collinear gluon density $xg(x,Q^2)$ is compared with that of global data fits NNPDF3.1sx  \cite{36} and CT14 \cite{53}. Later we investigate the sensitivity of the BFKL intercept in both $x$ and $k_t^2$ evolution of ugd's $f(x,k_t^2)$. Finally in Sect.~\ref{con} we summarize and outline our conclusion as  well as possible future prospects.
	
	\section{Theory}\label{theory}
	Recall that BFKL equation can be written as an integral equation for unintegrated gluon distributions $f(x,k_t^2)$ in the form\cite{12},
	\begin{equation}
	\label{3}
	f(x,k_t^2)=f_0(x,k_t^2) + \frac{\alpha_s}{2\pi}\int_{x}^{1}\frac{dz}{z}\int_{k_{0}^{'2}}d k_t^{'2}\kappa(k^2,k_t^{'2},z)f\left(\frac{x}{z},k_t^{'2}\right),
	\end{equation}
	where the inhomogeneous driving term $f^{(0)}(x,k_T^2)$ depicts gluon-proton coupling. The BFKL kernel is evaluated as 
	\begin{equation}
	\label{4}
	\kappa(k^2,k_t^{'2},z)f\left(\frac{x}{z},k_t^{'2}\right)=2N_c\frac{k_t^2}{k_t^{'2}}\left[\frac{\Theta \left(\frac{k_t^2}{z}-k_t^{'2}\right)f\left(\frac{x}{z},k_t^{'2}\right)-f\left(\frac{x}{z},k_t^2\right)}{\left| k_t^{'2}-k_t^2\right| }+\frac{f\left(\frac{x}{z},k_t^2\right)}{\sqrt{k_t^4+4k_t^{'4}}}\right].
	\end{equation}
	So called kinematic constraint $k_t^{'2}<k^2/z$ is imposed onto the real-emission part of the kernel \eqref{4} through the heaviside theta function $\Theta (\frac{k_t^2}{z}-k_t^{'2})$. The BFKL kernel that we incorporated in \eqref{4} is at leading logarithmic $1/x$ (LLx) accuracy, whereas higher order correction upto NLLx are found to be quite large \cite{2}. Implementation of the constraint in the evolution ensures the participation of only the LLx part of the higher order correction making the theory more realistic, which in fact, highlights the importance of NLLx correction.
	\par
	To get an integro differential equation, we differentiate \eqref{4} w.r.t. $\ln(1/x)$ and then using properties of $\Theta$ and Dirac-$\delta$ function viz.  $\Theta '(t)=\delta  (t)$ and $f( t)\delta  (t-a)=f(a)\delta  (t-a)$, one may show that
	
	\begin{equation}
	\label{5}
	\begin{split}
	\frac{\partial}{\partial \ln\frac{1}{x}}\int_{x}^{1}\frac{\text{d}z}{z}\Theta \left(\frac{k_t^2}{z}-k_t^{'2}\right)f(x,k_t^{'2})\longrightarrow          \Theta(k_t^2-k_t^{'2})f(x,k_t^{'2})+\Theta(k_t^{'2}-k_t^{2})f\left(\frac{k_t^{'2}}{k_t^{2}}x,k_t^{'2}\right).
	\end{split}
	\end{equation}
	Considering \eqref{5}, we can express \eqref{4} in the following integro-differential form,
	\begin{align}
	\label{6}
	\begin{split}
	-x\frac{\partial f\left(x,k_t^2\right)}{\partial x}=&\frac{\alpha _s k_t^2 N_c }{\pi }{\int _{k_{0}^{'2}}}\frac{dk_t^{'2}}{k_t^{'2}}\bigg[\frac{\Theta \left(k_t^2-k_t^{'2}\right)f\left(x,k_t^{'2}\right)+\Theta (k_t^{'2}-k_t^{2})f\left(\frac{k_t^{'2}}{k_t^{2}}x,k_t^2\right)}{\left| k_t^{'2}-k_t^2\right| }\\&-\frac{f(x,k_t^2)}{|k_t^{'2}-k_t^2|}+\frac{f\left(x,k_t^2\right)}{\sqrt{k_t^4+4k_t^{'4}}}\bigg],
	\end{split}
	\end{align} 
	where we neglect the term $x\frac{\partial f_0}{\partial x}$ in \eqref{6} as it is much less singular than $x\frac{\partial f}{\partial x}$ at small $x$ \cite{6,15}.

	\par In pQCD the small-$x$ behavior of the parton distribution is modulated by the intercept of appropriate Regge trajectory. 
	The BFKL dynamics itself is based on the concept of pomeranchuk theorem or pomeron: the Regge-pole carrying the quantum-numbers of the vacuum.  The BFKL-pomeron intercept is yielded by
	\begin{equation*}
	\alpha_{P}^{BFKL}(0) = 1 + \lambda_{\text{BFKL}},
	\end{equation*}
	where  $\lambda_{\text{BFKL}}= \frac{3 \alpha_s}{\pi} 28  \zeta(3)$, $\zeta$ being Reiman zeta function \cite{9}. However, the BFKL hard pomeron should be contrasted with non perturbative description of soft pomeron in the sense that BFKL intercept is potentially large $\alpha_{P}^{BFKL}(0)\approx1.5$ (or $\lambda_{\text{BFKL}}\sim 0.5$) for $\alpha_s$=0.2 compared to that of soft pomeron $(\alpha_{P}(0)=1.08)$. 
	
	Regge model provides a steep power law parametrization of DIS distribution functions, $f_i(x,Q^2)=A_i(Q^2)x^{-\lambda_i}$ (i=$\sum$ (singlet structure function)and g (gluon distribution)) \cite{7,8}.  This motivates us to consider a simple form of Regge factorization  as follows,
	\begin{equation}
	\label{7}
	\begin{split}
	f\left(\frac{k_t^{'2}}{k_t^2}x,k_t^2\right)&\simeq x^{-\lambda_{\text{BFKL}}}\left(\frac{k_t^{2}}{k_t^{'2}}\right)^{\lambda_{\text{BFKL}}} \mathcal{L}(k_t^2)= \left(\frac{k_t^2}{k_t^{'2}}\right)^{\lambda }f\left(x,k_t^2\right),
	\end{split}
	\end{equation}
	where we drop the subscript on $\lambda_{\text{BFKL}}$ (referred to as $\lambda$ henceforth). Note that Regge factorization can't be considered as good ansatz for the entire kinematic domain of $x$ and $k_t^2$ \cite{80} and it is supposed to be valid only if the quantity invariant mass, $W$ $(=\sqrt{Q^2(1-x)/x})$ is much greater than all other variables. Therefore, we expect the Regge factoriztion in \eqref{7} to be justified if $x$ is small enough, for any value of $Q^2$.

	Now considering the Regge factorization \eqref{7} and taking $\Theta(\omega)=1-\Theta(-\omega)$ we can express \eqref{6} as
	\begin{equation}
	\label{8}
	\begin{split}
	-x\frac{\partial f\left(x,k_t^2\right)}{\partial x}=&\frac{\alpha _s N_c k_t^2 }{\pi }\int _{k_0^{'2}}\frac{dk'^2}{k_t^{'}{}^2}\bigg[\Theta \left(k_t^2-k_t^{'2}\right)\frac{1-\left(\frac{k_t^2}{k_t^{'2}}\right)^{\lambda }}{\left| k_t^{'2}-k_t^2\right| }f\left(x,k_t^{'2}\right)\\
	&+\left(\frac{k_t^2}{k_t^{'2}}\right)^{\lambda }\frac{f\left(x,k_t^2\right)}{\left| k_t^{'2}-k^2\right|}-\frac{f\left(x,k_t^2\right)}{\left| k_t^{'2}-k^2\right|} +\frac{f\left(x,k_t^2\right)}{\sqrt{k_t^4+4k_t^{'4}}}\bigg].
	\end{split}
	\end{equation}
	The above equation \eqref{8} is the integral form of kinematic constraint improved BFKL equation.
	\par As we have mentioned earlier, transverse gluon momenta in BFKL multi-Regge kinematics pose no ordering and this allows us to write the gluon distribution in Taylor series,
	\begin{equation}
	\label{9}
	f\left(x,k_t^{'2}\right)=f\left(x,k_t^2\right)+\frac{\partial f\left(x,k_t^2\right)}{\partial k_t^2}\left(k_t^{'2}-k_t^2\right) +  \mathcal{O}\left(k_t^{'2}-k_t^2\right),
	\end{equation}
	where the higher order terms in the series are denoted by $\mathcal{O}\left(k_t^{'2}-k_t^2\right)$. The convergence of the series is inherent because of the no ordering of the transverse momenta $k_t^{'}- k_t\simeq 0$ implising higher order terms  $\mathcal{O}\left(k_t^{'2}-k_t^2\right)\rightarrow 0$. This ensures the insignificance of the higher order terms and thereby $\mathcal{O}\left(k_t^{'2}-k_t^2\right)\rightarrow 0$  can be neglected. Thus this assumption would hold good as long as no ordering condition of the transverse momenta in BFKL kinematics is concerned.  This type of series expansion of unintegrated gluon distribution is well supported in the literature \cite{12}.
	
	\par Now neglecting these higher order terms  we can express \eqref{8} as
	\begin{equation}
	\label{10}
	\begin{split}
	-x\frac{\partial f\left(x,k_t^2\right)}{\partial x}=\xi(k_t^2) \frac{\partial f\left(x,k_t^2\right)}{\partial k_t^2}
	+\zeta(k_t^2) f\left(x,k_t^2\right),
	\end{split}
	\end{equation}
	where
	\begin{equation}
	\label{11}
	\begin{split}
	\xi(k_t^2)=\frac{\alpha _s N_c }{\pi }k_t^2\bigg[\int _{k_t^{2}}^\infty\frac{dk_t^{'2}}{k_t^{'2}}\left(\frac{k_t^2}{k_t^{'2}}\right)^{\lambda }\frac{k_t^{'2}-k_t^2}{\left| k_t^{'2}-k_t^2\right| }+\int _{k_0^{'2}}^{k_t^2}\frac{dk_t^{'2}}{k_t^{'2}}\frac{k_t^{'2}-k_t^2}{\left| k_t^{'2}-k_t^2\right| }\bigg],
	\end{split}
	\end{equation}
	
	\begin{equation}
	\label{12}
	\begin{split}
	\zeta(k_t^2)=\frac{\alpha _s N_c }{\pi }k_t^2\bigg[\int _{k_t^2}^\infty\frac{dk_t^{'2}}{k_t^{'2}}\left(\frac{k_t^2}{k_t^{'2}}\right)^{\lambda }\frac{1}{\left| k_t^{'2}-k_t^2\right| }+\int _{k_0^{'2}}^{\infty }\frac{dk_t^{'2}}{k_t^{'2}}\frac{1}{\sqrt{k_T^4+4k_t^{'4}}}-\int _{k_t^2}^{\infty }\frac{dk_t^{'2}}{k_t^{'2}}\frac{1}{\left| k_t^{'2}-k_t^2\right| }\bigg].
	\end{split}
	\end{equation}
	We get rid of the infrared singularities in the integrals $I_1=\int _{k_t^2}^{\infty }\frac{dk_t^{'2}}{k_t^{'2}}\frac{1}{\left| k_t^{'2}-k_t^2\right| }$ and $I_2=\int _{k_t^2}\frac{dk_t^{'2}}{k_t^{'2}}\left(\frac{k_t^2}{k_t^{'2}}\right)^{\lambda }\frac{1}{\left| k_t^{'2}-k_t^2\right| }$ performing some angular integral prescription. First we consider the following two fold azimuthal integral with the replacement of the variable $k_t^{'} \rightarrow k_t^{'} +k_t$ on the r.h.s.,
	\begin{equation}
	\label{14}
	\int\frac{\text{d}^2k_t^{'}}{(k_t^{'}-k_t)^2}\frac{1}{(k_t^{'}-k_t)^2+k_t^{'2}}=\int\frac{\text{d}^2k_t^{'}}{k_t^{'2}}\frac{1}{k_t^{'2}+(k_t^{'}+k_t)^2}.
	\end{equation}
	Now recalling the standard trigonometric integral
	\begin{equation}
	\label{13}
	\int_{0}^{2\pi}\frac{\text{d}\theta}{P+Q\cos\theta}=\frac{2\pi}{\sqrt{P^2+Q^2}},
	\end{equation} 
	we can express the l.h.s. and r.h.s. of \eqref{14} as follows
	\begin{align}
	\label{16}
	\int\frac{\text{d}^2k_t^{'}}{(k_t^{'}-k_t)^2}\frac{1}{(k_t^{'}-k_t)^2+k_t^{'2}} =\pi \int \frac{dk_t^{'2}}{k_t^{'2}}\frac{1}{\left| k_t^{'2}-k_t^2\right| }-\pi\int\frac{\text{d}k_t^{'2}}{k_t^{'2} \sqrt{4k_t^{'4}+k_t^4}},\\
	\label{15}
	\int\frac{\text{d}k_t^{'}  k_t^{'} \text{d}\theta}{k_t^{'2}(2k_t^{'2}+k_t^2+2k_t^{'}k_t\cos\theta)}=2\pi\int\frac{\text{d}k_t^{'} k_t^{'}}{k_t^{'2} \sqrt{4k_t^{'4}+k_t^4}}=\pi\int\frac{\text{d}k_t^{'2}}{k_t^{'2} \sqrt{4k_t^{'4}+k_t^4}}.
	\end{align}
	From \eqref{16} and \eqref{15} we obtain,
	\begin{equation}
	\label{17}
	\int\frac{dk_t^{'2}}{k_t^{'2}}\frac{1}{\left| k_t^{'2}-k_t^2\right| }= 2 \int\frac{\text{d}k_t^{'2} }{k_t^{'2} \sqrt{4k_t^{'4}+k_t^4}},
	\end{equation}
	which turns out to be well-behaved for both of the integration limits in the integral $I_1$. For moderate  $k_t^{'2}$, in particular at $k_{T_{\text{min}}}^{'2} \leq k_t^{'2}\leq k_t^2$, the longitudinal contribution to the gluon virtuality is negligible, preserving the no ordering condition of transverse momenta $k_t^{'2}\approx k_t^2$ very strictly. In order to make our calculation simple, without altering the underlying physics, we therefore implant a factor $\left(k_t^{2}/k_t^{'2}\right)^\lambda$ inside the integrals, particularly with integration limit $k_{T_{\text{min}}}^{'2} \leq k_t^{'2}\leq k_t^2$. In addition, this removes the infrared divergence  of the second improper integral $I_2$ lowering the infrared limit down to $k_0^{'2}$.

	\par Now considering these approximations in \eqref{11} and \eqref{12} and putting $I_1$ in \eqref{12} we obtain
	\begin{align}
	\label{18}
	\begin{split}
	\xi(k_t^2)=&\frac{\alpha _s N_c k_t^2}{\pi }\bigg[\int _{k_{0}^{'2}}^{k_t^2}\frac{dk_t^{'2}}{k_t^{'2}}\left(\frac{k_t^2}{k_t^{'2}}\right)^\lambda\frac{k_t^{'2}-k_t^2}{\left| k_t^{'2}-k_t^2\right| }+\int _{k_t^{2}}^{\infty}\frac{dk_t^{'2}}{k_t^{'2}}\left(\frac{k_t^2}{k_t^{'2}}\right)^{\lambda }\frac{k_t^{'2}-k_t^2}{\left| k_t^{'2}-k_t^2\right| }\bigg]\\=&\frac{\alpha_s N_c k_t^2}{\pi}\frac{1}{\lambda}\left(2-k_t^{2\lambda}\right)\approx-\frac{\alpha_s N_c}{\pi\lambda}(k_t^2)^{\lambda+1},
	\end{split}
	\end{align}
	\begin{align}
	\label{19}
	\begin{split}
	\zeta(k_t^2)=&\frac{\alpha _s N_c }{\pi }k_t^2\bigg[\int _{k_{0}^{'2}}^{\infty }\left(\frac{k_t^2}{k_t^{'2}}\right)^\lambda\frac{dk_t^{'2}}{k_t^{'2}}\frac{1}{\left| k_t^{'2}-k_t^2\right| }
	-\int _{k_{0}^{'2}}^{\infty }\frac{dk_t^{'2}}{k_t^{'2}}\frac{1}{\sqrt{k_t^4+4k_t^{'4}}}\bigg]\\
	=&\frac{\alpha _s N_c }{\pi }\bigg[\frac{k_t^{2\lambda}}{\lambda}- 2^{2-\frac{\lambda }{2}}  \lambda^{-1}k_t^{2\lambda}\bigg(1-\frac{\sqrt{k_t^4+4}}{k_t^2}\bigg)^{\lambda /2} \, _2F_1-\lambda\ln \left(\frac{k_t^2}{2}+\sqrt{1+\frac{k_t^4}{4}}\right)\bigg]
	\approx \frac{\alpha _s N_c }{\pi }\left(\epsilon+\frac{(k_t^2)^\lambda }{\lambda}\right),
	\end{split}
	\end{align}
	where $\, _2F_1$=$\, _2\text{F}_1\left(-\frac{\lambda }{2},\frac{\lambda }{2};1-\frac{\lambda }{2};\frac{k_t^2+\sqrt{k_t^4+4}}{2 k_t^2}\right)$ is a standard hypergeometric function. We have chosen infrared cutoff $k_{0}^{'2}= 1 \text{ GeV}^2$ as this suggested a very consistent results towards HERA DIS  data for proton structure function $F_2$ in the ref.~\cite{14}.  From phenomenological studies we have found that the term with hypergeometric function in \eqref{19}   becomes irrelevant towards change in $k_t^2$ and tend to a constant value $(\approx -4.79)$. For small enough $k_t^2$, this contribution can not be overlooked since at this range, the $k_t^{2\lambda}/\lambda$ contribution  itself is small. On the other hand,  the logarithmic term contribution in \eqref{19} is negligible in comparison to the net contribution for the entire $k_t^2$ domain of study. In consequence, the constant contribution from the hypergeometric term can be treated as a small perturbation $\epsilon$ to the dominant term $k_t^{2\lambda}/\lambda$ . We have performed nonlinear regression technique to reduce the error to a minimum in determination of the constant perturbation parameter $\epsilon$.
	
	\par
\begin{figure*}[t]
		\centering 
		\includegraphics[width=.45\textwidth]{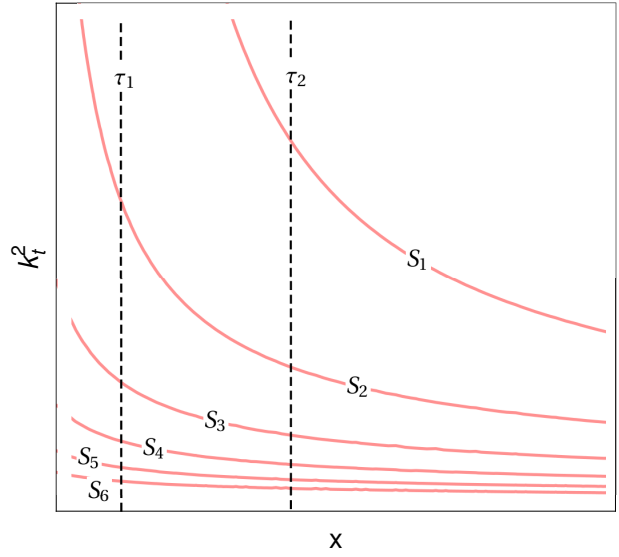}
		
		\caption{\label{f2} Characteristic curves obtained from the characteristic equations \eqref{20} and \eqref{21} in $x$-$k_t^2$ plane.}
	\end{figure*}

 Now we are set to solve the PDE \eqref{10}. To introduce the method of characteristics, let us recast the variables $x$ and $t$ in terms of two new variables $S$ and $\tau$, such that
	\begin{align}
	\label{20}
	&\frac{d x}{d S} = x,\\
	&\label{21}
	\frac{d k_t^2}{d S}=\xi(k_t^2),
	\end{align}
	which are known as characteristic equations. Fig.~\ref{f2} indicates characteristic curves in the $x$-$k_t^2$ plane. Now putting \eqref{20} and \eqref{21} in \eqref{10} we get
	\begin{align}
	\label{22}
	\frac{d f(S,\tau)}{d S}+\zeta (S, \tau)/2f(S,\tau)=0,
	\end{align}
	which can be solved as
	\begin{align}
	\label{23}
	f(S,\tau)= f(0,\tau)e^{-\int_{0}^{S}\zeta\left(S,\tau\right)d S}.
	\end{align}
	
	For evolution of $x$ the gluon distribution function varies with $x$, while $k_t^2$ remains constant. Hence \eqref{20} can be used in \eqref{22}, the solution of which is yielded by
	\begin{align}
	\label{24}
	f(S,\tau)= f(\tau)\left(\frac{x}{x_0}\right)^{\zeta(S, \tau)/2},
	\end{align}
	where  $f(S,\tau)= f(\tau)$ for $S=0$, $x=x_0$, provided $x_0$ is chosen small enough to ensure the validity of  BFKL equation.
	\par Replacing the coordinate system ($S,\tau$) to its original one ($x,k_t^2$), we get the $x$ evolution of the unintegrated gluon distribution as
	\begin{align}
	\label{25}
	f(x,k_t^2)= f(x_0,k_t^2)\left(\frac{x}{x_0}\right)^{\zeta(k_t^2)/2}.
	\end{align}
	Similarly the $k_t^2$ evolution of the unintegrated gluon distribution will be
	\begin{align}
	\label{26}
	f(x,k_t^2)= f(x,k_{0}^2)e^{-\int_{k_0^2}^{k_t^2}\frac{\zeta\left(k_t^2\right)}{\xi\left(k_t^2\right)}d k_t^2}.
	\end{align}
	In the following section, we present an analysis on the phenomenological aspect of the solutions for $x$ evolution \eqref{25} and $k_t^2$ evolution \eqref{26} by picking some appropriate initial boundary condition.
	
	\section{Results and discussion}\label{result}
	\begin{figure*}[t]
		\centering 
		\includegraphics[trim=25 5 21.5 0,width=.46\textwidth,clip]{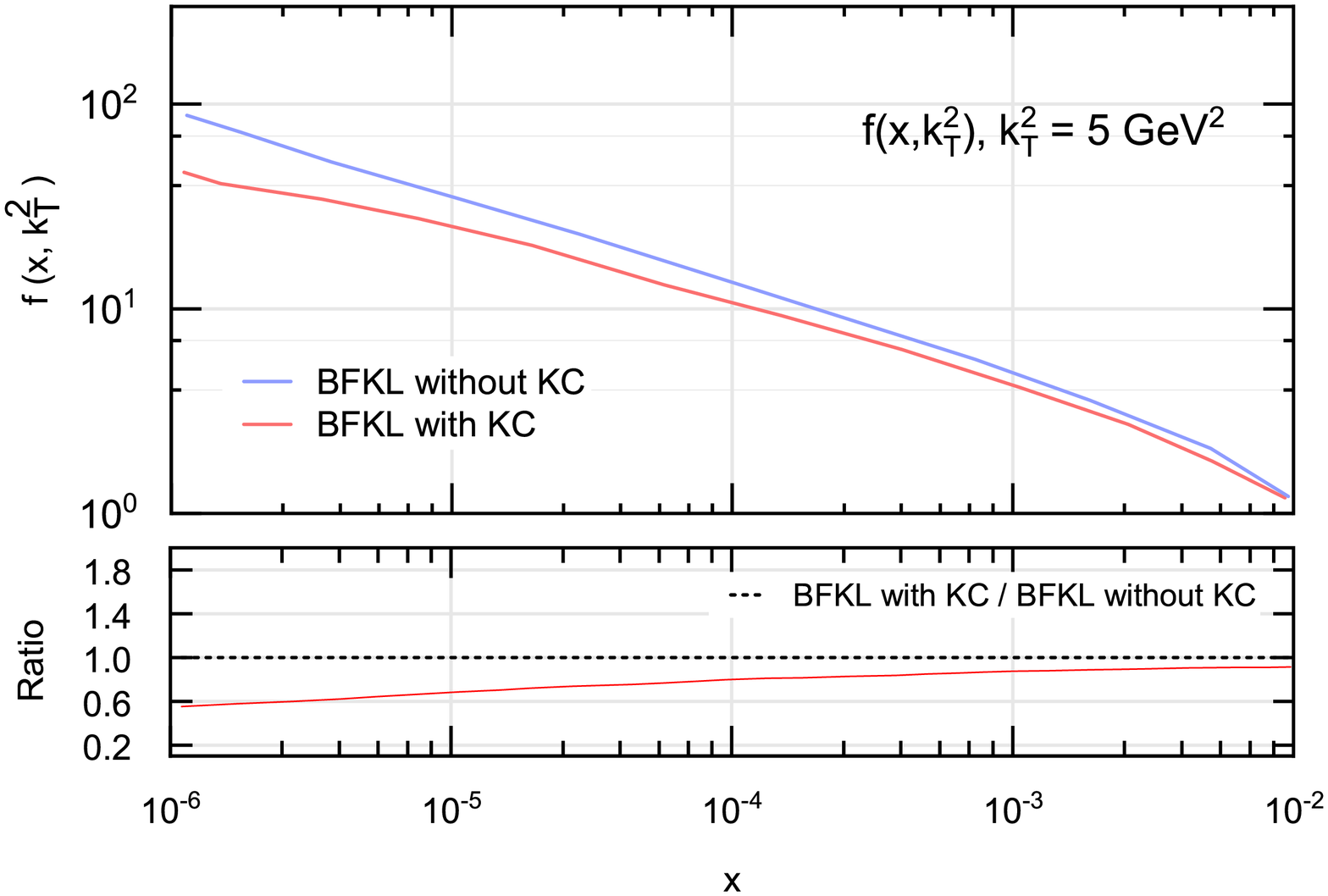}\hspace{5mm}
		\includegraphics[trim=25 5 21.5 0,width=.46\textwidth,clip]{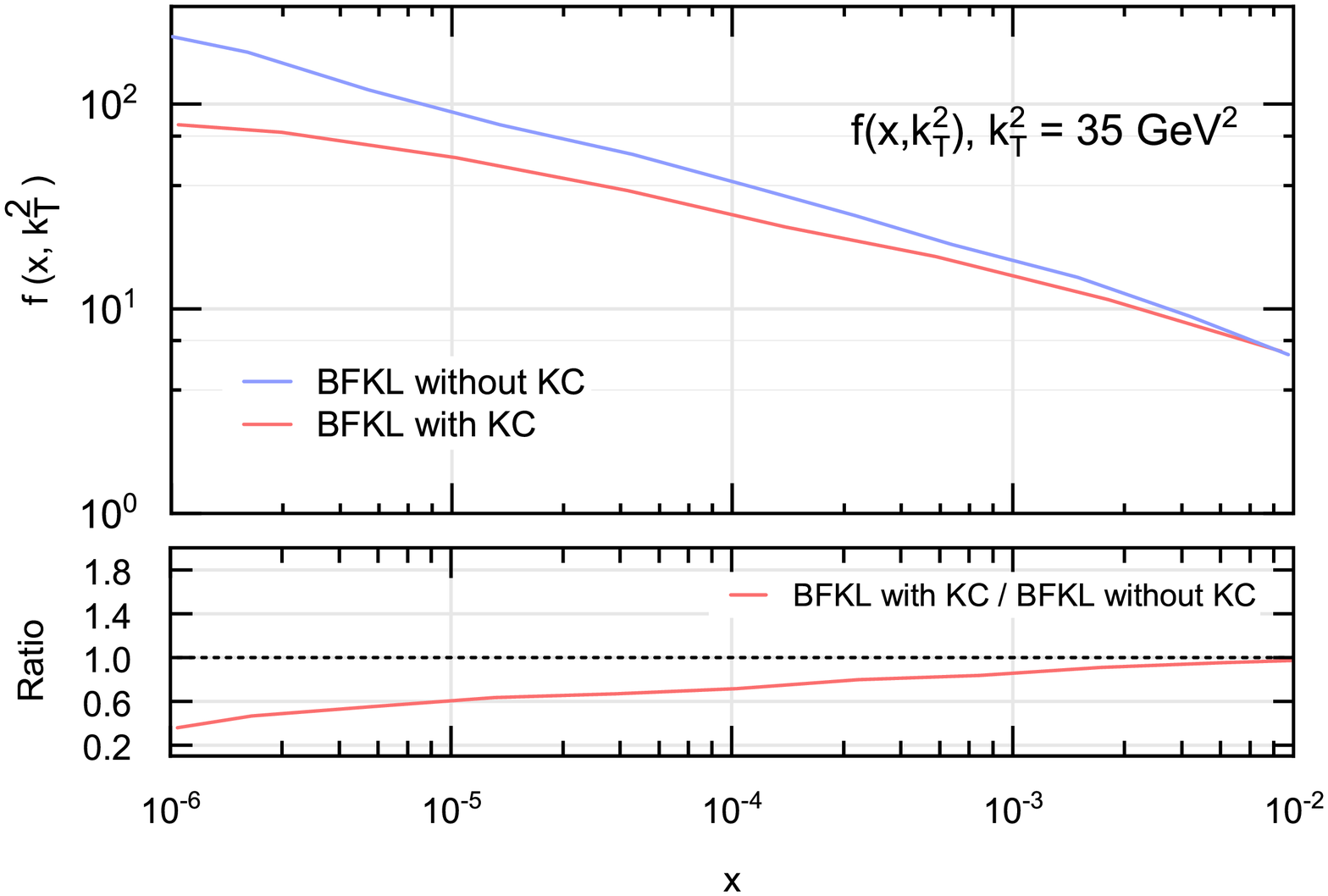}
		\includegraphics[trim=25 5 21.5 0,width=.46\textwidth,clip]{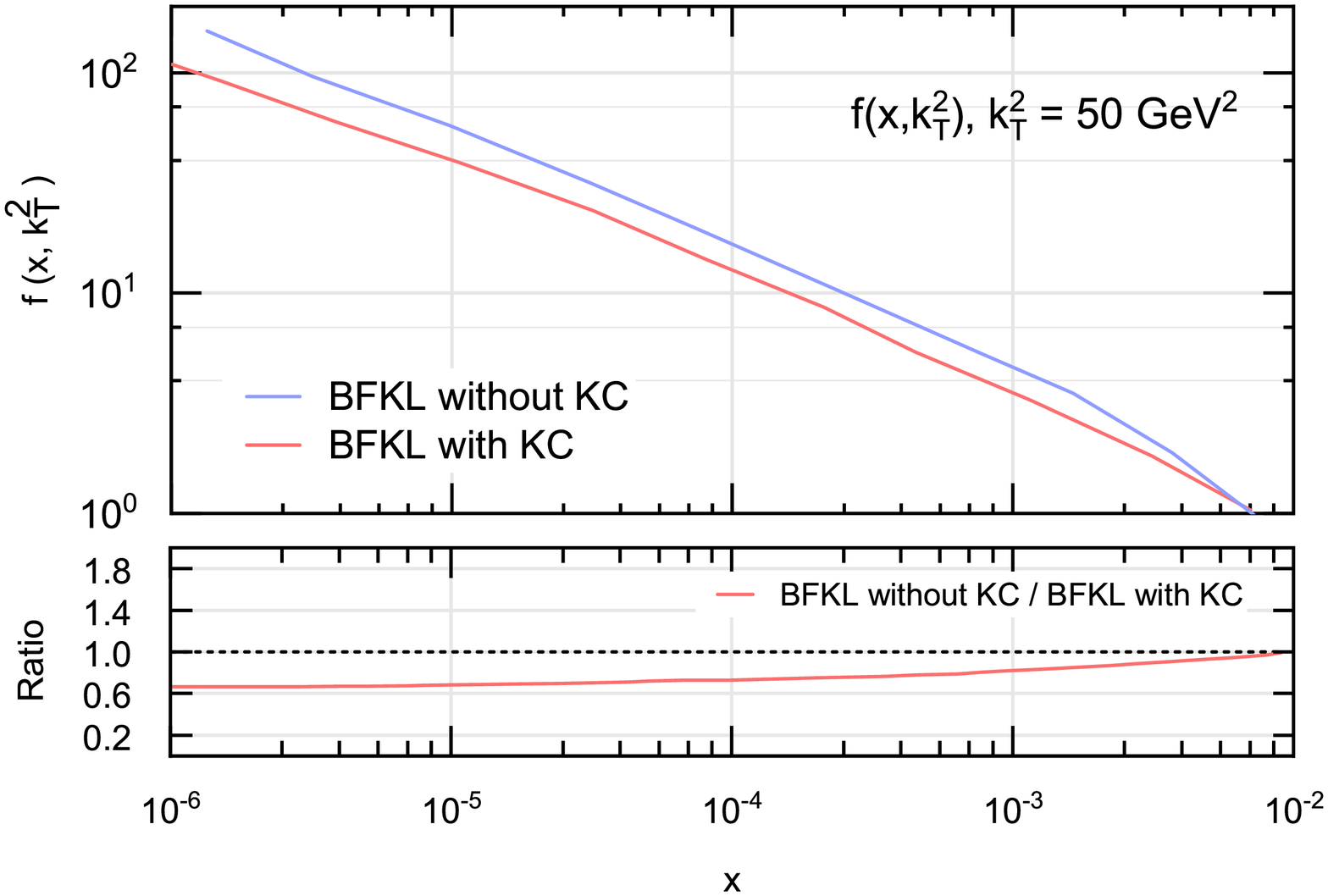}\hspace{5mm}
		\includegraphics[trim=25 5 21.5 0,width=.46\textwidth,clip]{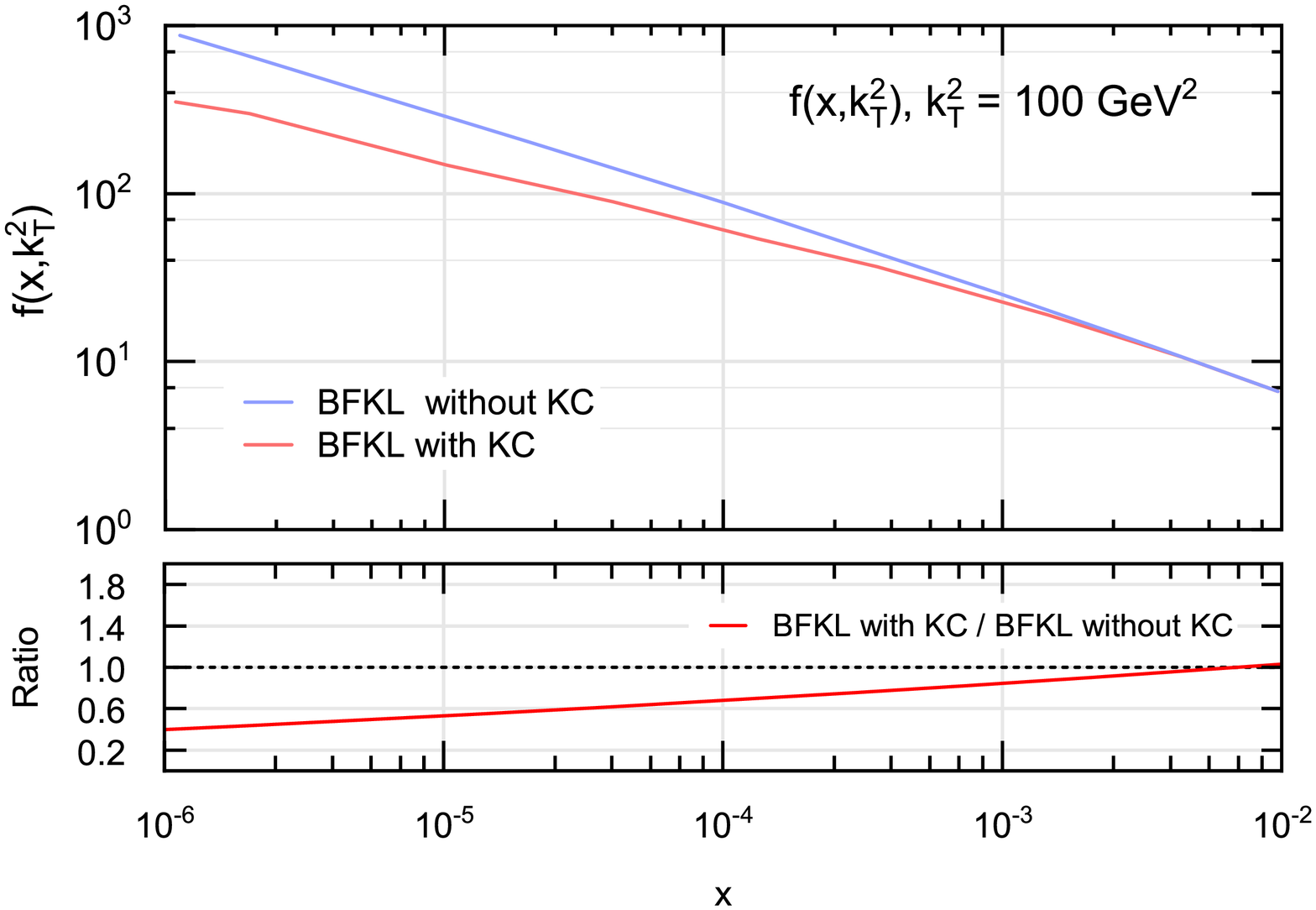}
		
		\caption{\label{f3}$x$ evolution of unintegrated gluon distribution $f(x,k_T^2)$. Our result of KC-BFKL evolution is contrasted with that of original BFKL evolution.}
	\end{figure*}
	
	\par
\begin{figure*}[t]
		\centering 
		\includegraphics[trim=25 5 21.5 0,width=.46\textwidth,clip]{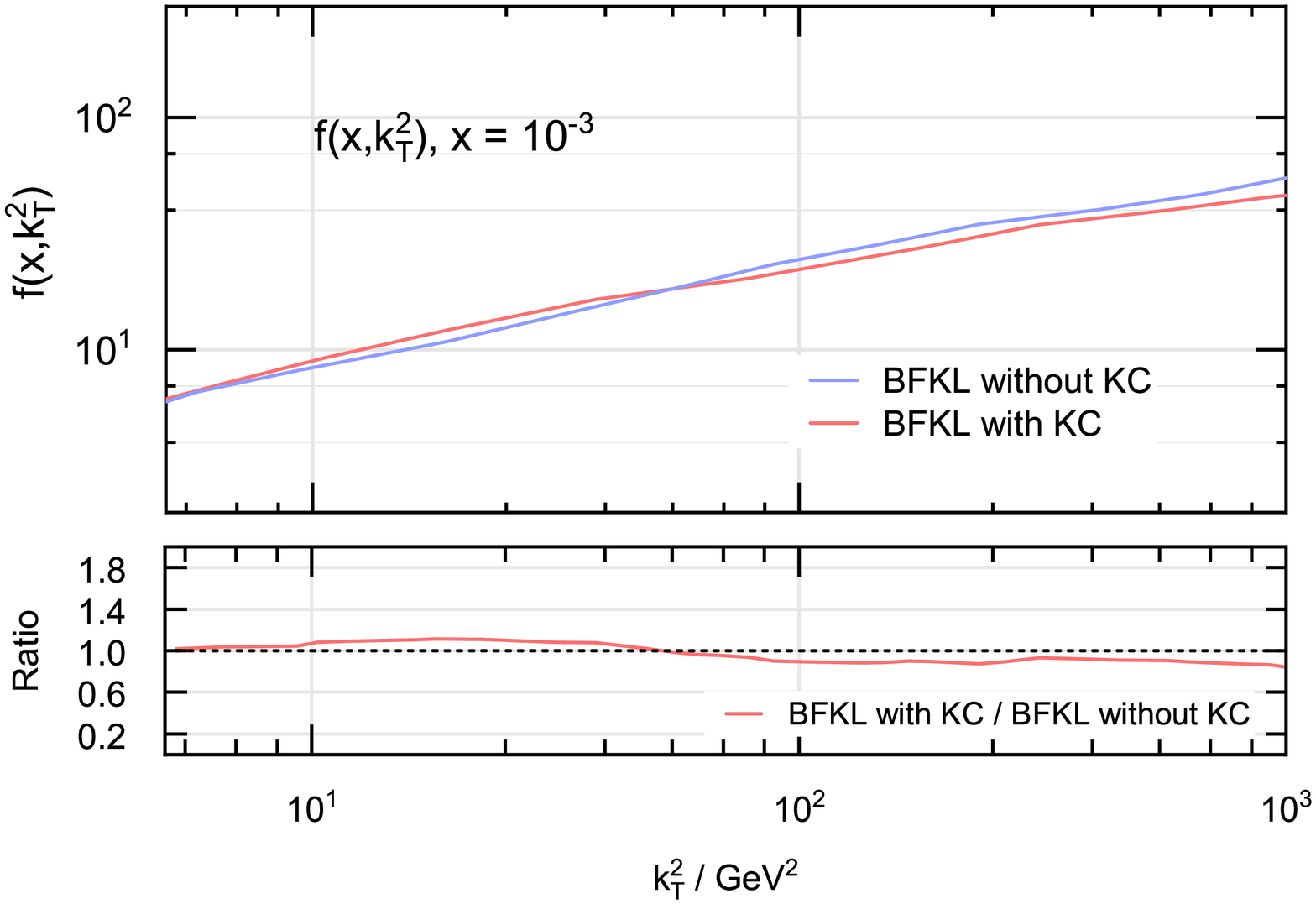}\hspace{5mm}
		\includegraphics[trim=25 5 21.5 0,width=.46\textwidth,clip]{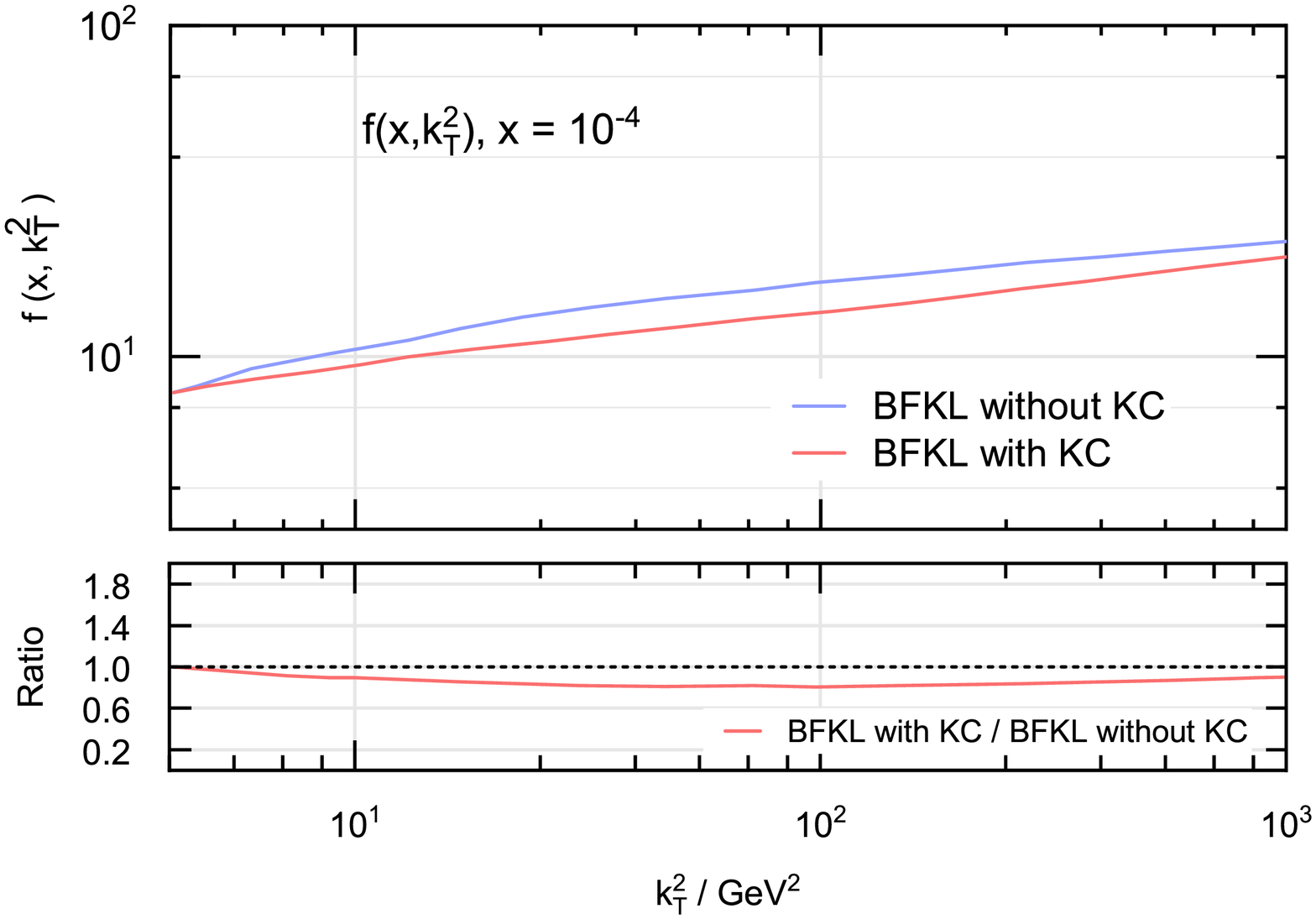}
		\includegraphics[trim=25 5 21.5 0,width=.46\textwidth,clip]{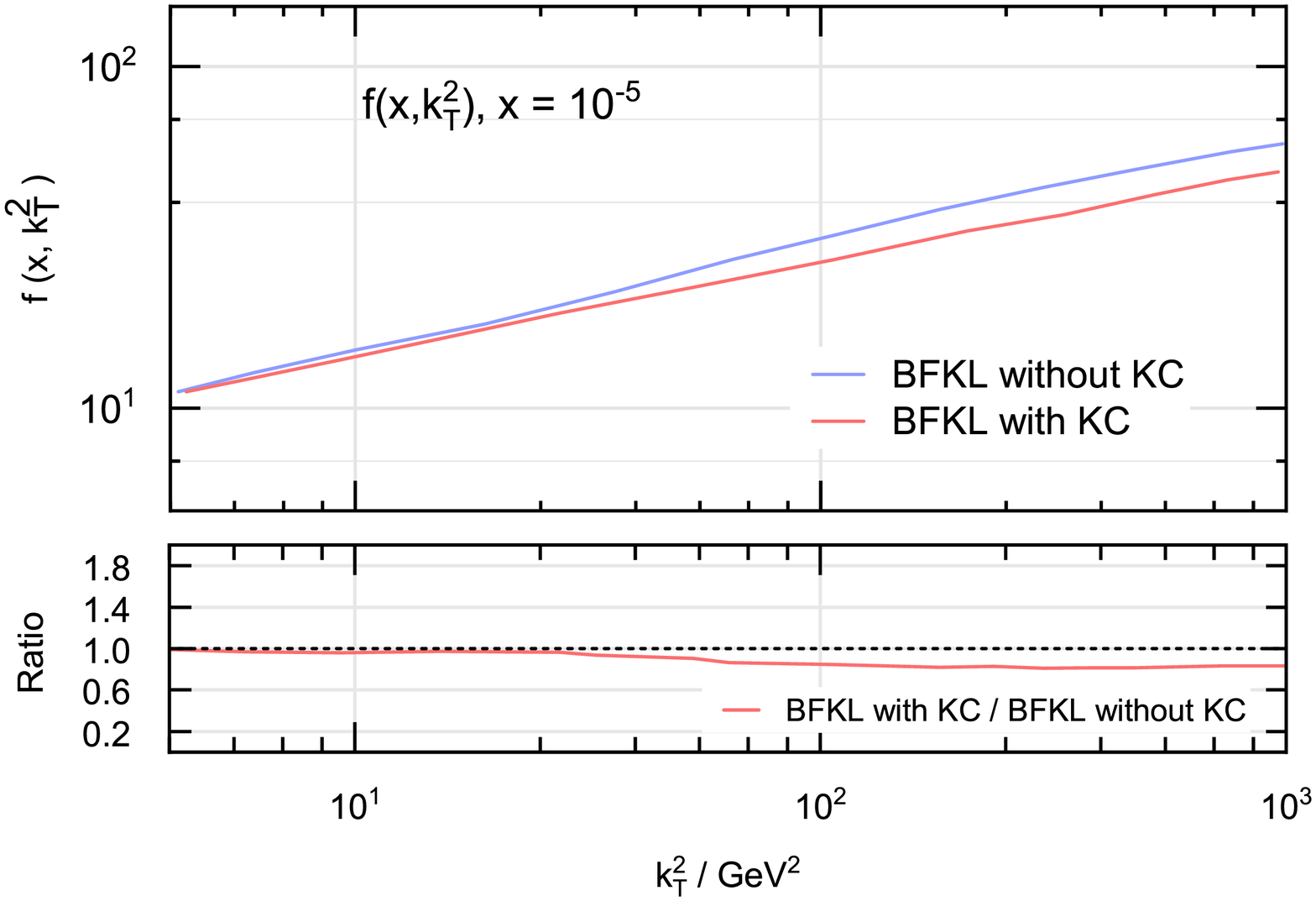}\hspace{5mm}
		\includegraphics[trim=25 5 21.5 0,width=.46\textwidth,clip]{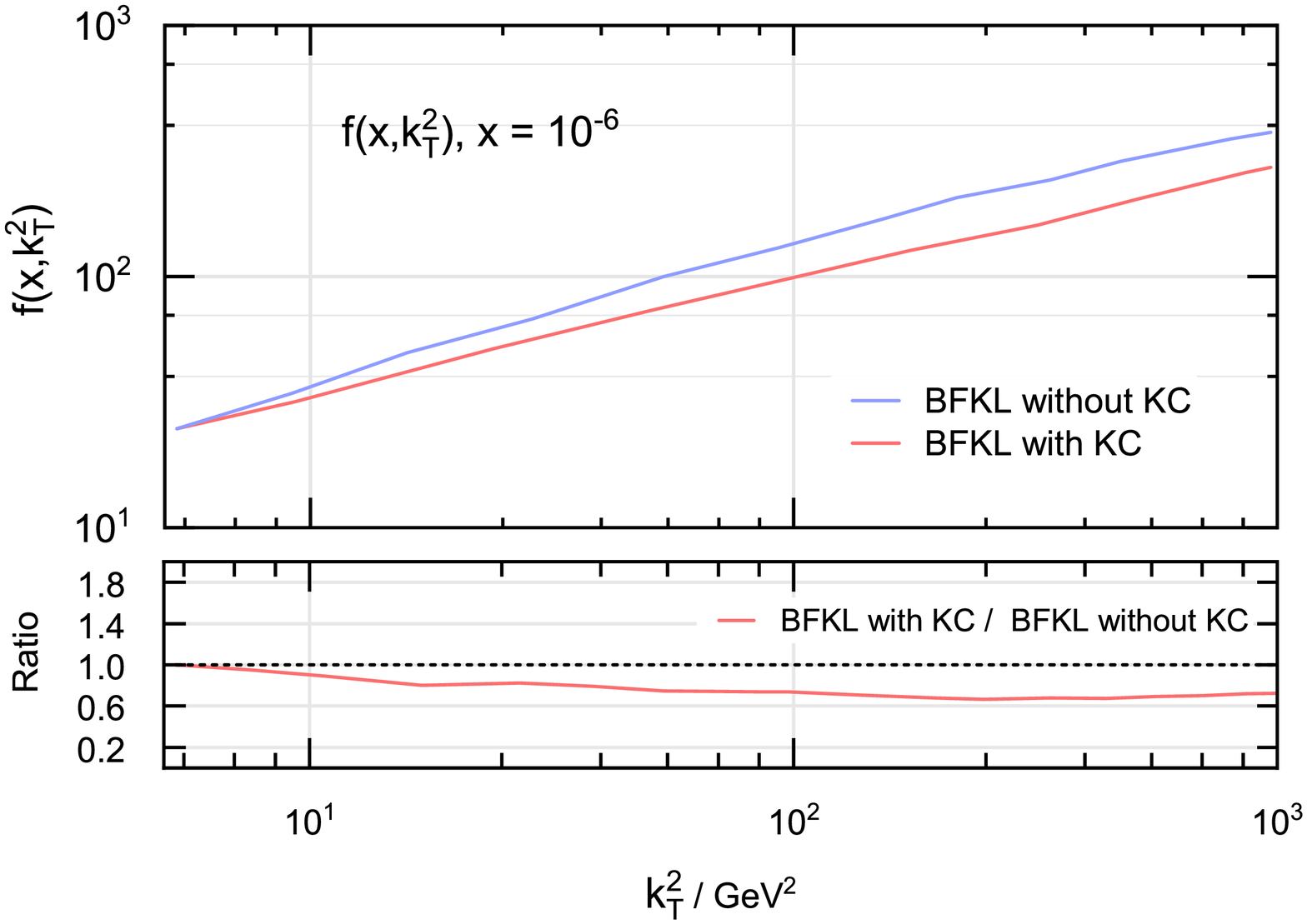}
		
		\caption{\label{f4}$k_t^2$ evolution of unintegrated gluon distribution $f(x,k_t^2)$. Our result of KC-BFKL evolution is contrasted with that of original BFKL evolution.}
	\end{figure*}

 Recall the well-known solution of the linear BFKL equation \cite{9}
	\begin{equation}
	\label{32}
	f(x,k_t^2)=\beta \frac{x^{-\lambda}\sqrt{k_t^2}}{\sqrt{\ln\frac{1}{x}}}\exp \left(-\frac{\ln^2(k_t^2/k_s^2)}{2\Omega\ln (1/x)}\right),
	\end{equation}
	where $\lambda= \frac{3 \alpha_s}{\pi} 28  \zeta(3)$, $\zeta$ being Riemann zeta function and $\Omega=32.1 \alpha_s$. The nonperturbative parameter $k_s^2= 1 \text{ GeV}^2$ and the normalization constant $\beta\sim 0.01$ \cite{1}. We take \eqref{32} as the input distribution for our solution of $x$ and $k_t^2$ evolution of KC-BFKL equation (equation \eqref{25} and \eqref{26} respectively). We have plotted the solution for respective $x$ and $k_t^2$ evolution in Fig.~\ref{f3} and Fig.~\ref{f4}. Our KC improved BFKL prediction is contrasted with ordinary BFKL evolution \eqref{32} to assess the effect of kinematic constraint in gluon evolution. The $x$ evolution of $f(x,k_t^2)$ is shown in Fig.~\ref{f3} for four $k_t^2$ values viz. $5 \text{ GeV}^2$, $35 \text{ GeV}^2$, $50 \text{ GeV}^2$ and $100 \text{ GeV}^2$.
	The input is taken at higher $x$ value $x=10^{-2}$ and then evolved down to smaller $x$ value upto $x=10^{-6}$ thereby setting the kinematic range of evolution $10^{-6}\leq x\leq10^{-2}$. We observe the singular $x^{-\lambda}$ growth of the gluon distribution in both of the evolutions, however the KC-BFKL is found to rise slowly compared to ordinary BFKL evolution towards very small $x$. The suppression in KC-BFKL compared to ordinary BFKL is roughly around 10-30\% at $x=10^{-6}$ for all $k_t^2$ bins, however it is hard to establish any significant distinction between the two for $x\geq 10^{-3}$ regime. The $k_t^2$ evolution is studied for the kinematic range $5 \text{ GeV}^2\leq k_t^2\leq 10^3 \text{ GeV}^2$ corresponding to four different values of $x$ as indicated in Fig.~\ref{f4}. Both KC-BFKL and ordinary BFKL forecast similar kind of growth, however the evolution is  slightly suppressed in case of KC-BFKL. At smaller $x$ bins ($x=10^{-5}$, $10^{-6}$), the distinction between the two evolutions becomes more prominent.
	\par
\begin{figure*}[t]
		\centering 
		\includegraphics[trim=25 5 21.5 0,width=.46\textwidth,clip]{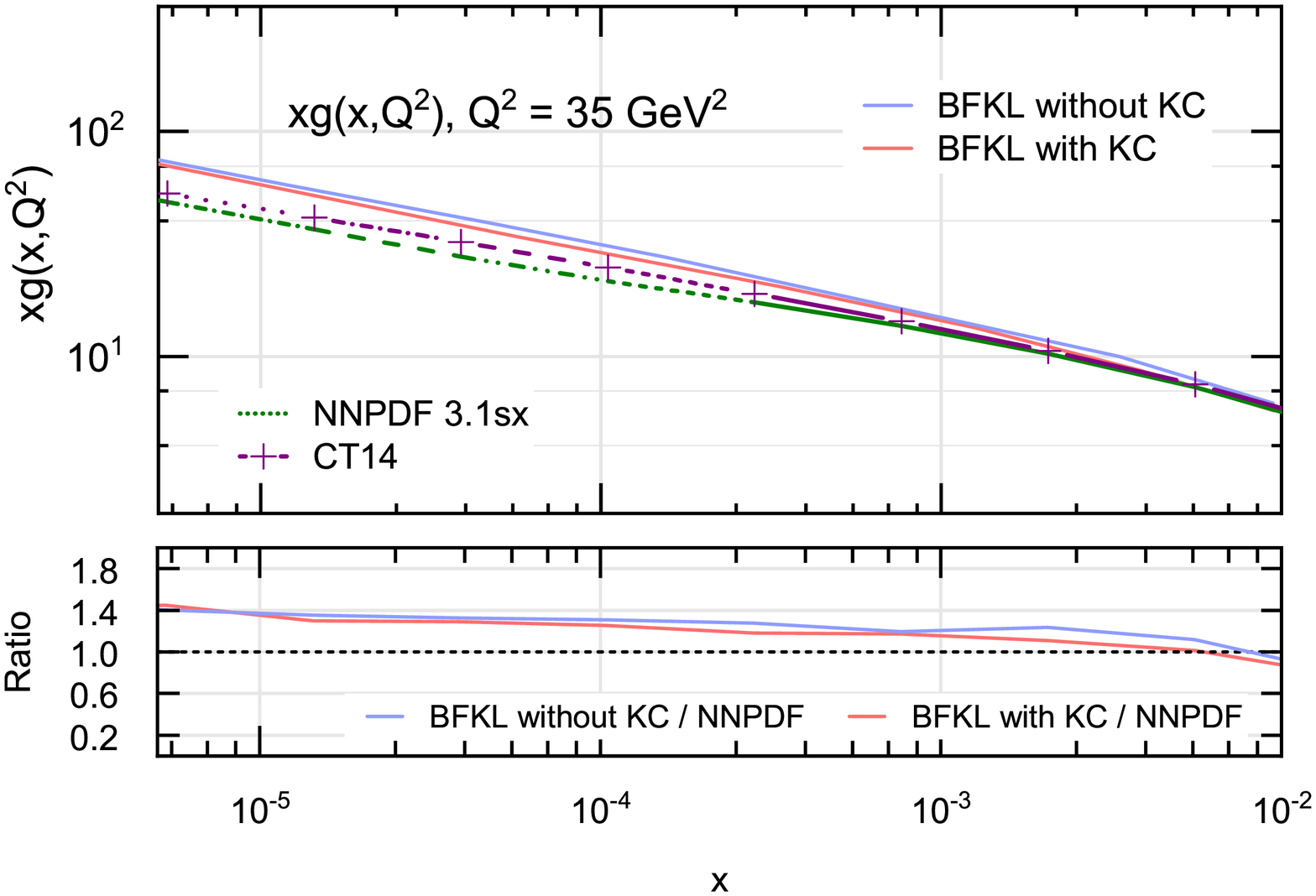}\hspace{5mm}
		\includegraphics[trim=25 5 21.5 0,width=.46\textwidth,clip]{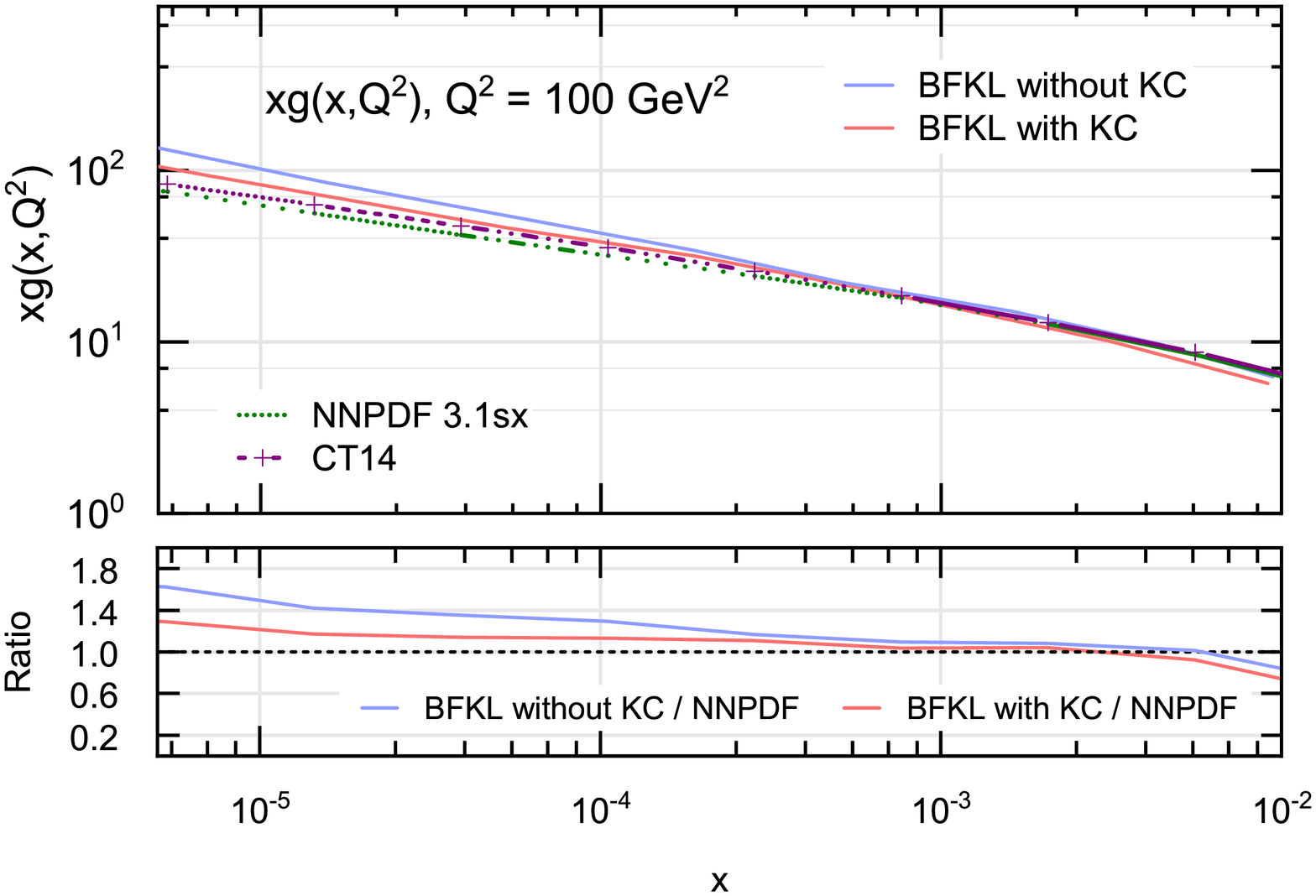}
		
		\caption{\label{f5}$x$ evolution of collinear gluon distribution $xg(x,Q^2)$. Our extracted collinear gluon density from KC-BFKL evolution is contrasted with that of global data fits NNPDF3.1sx and CT14.}
	\end{figure*}

\begin{figure*}[t]
		\centering 
		\includegraphics[trim=25 5 21.5 0,width=.46\textwidth,clip]{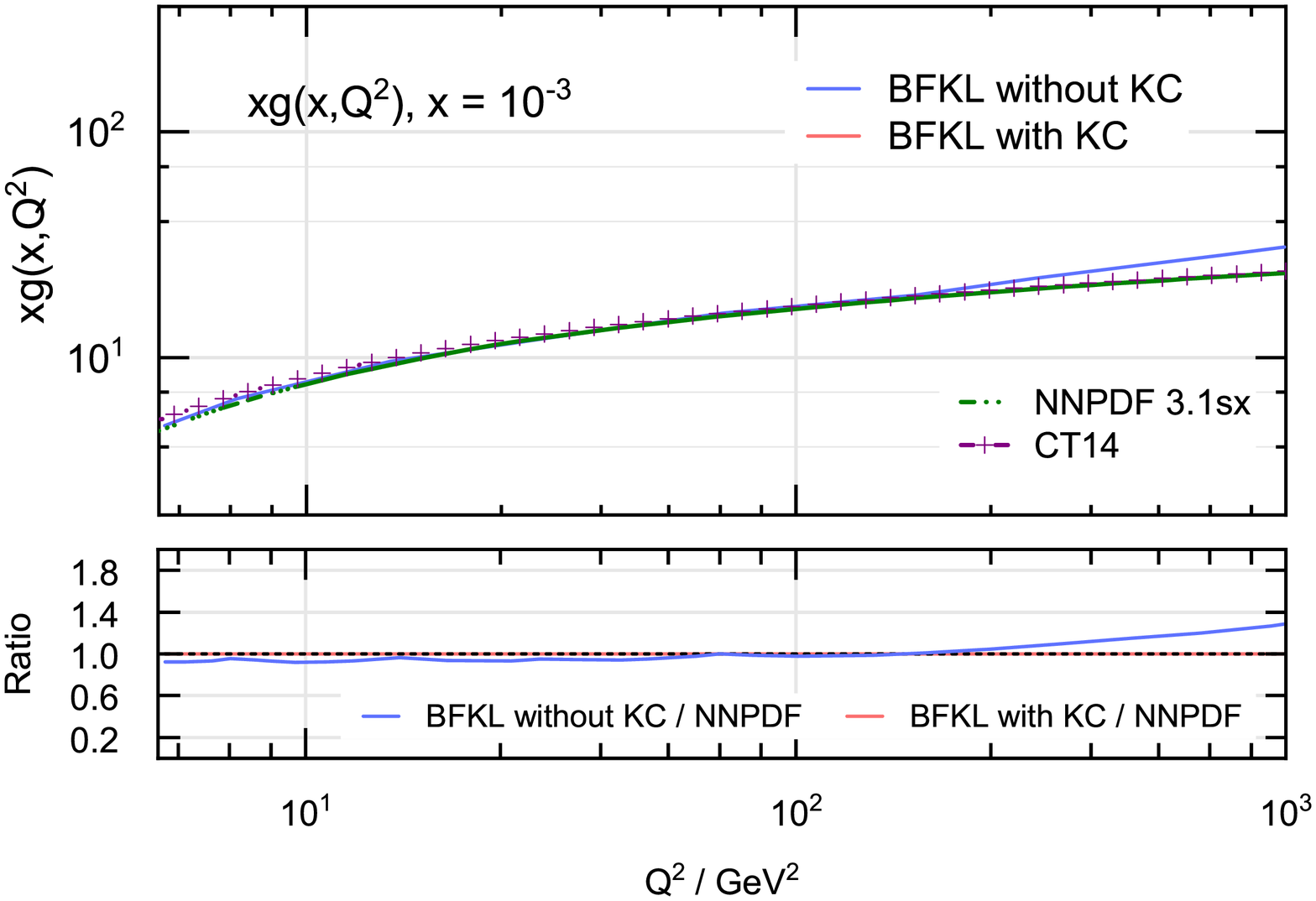}\hspace{5mm}
		\includegraphics[trim=25 5 21.5 0,width=.46\textwidth,clip]{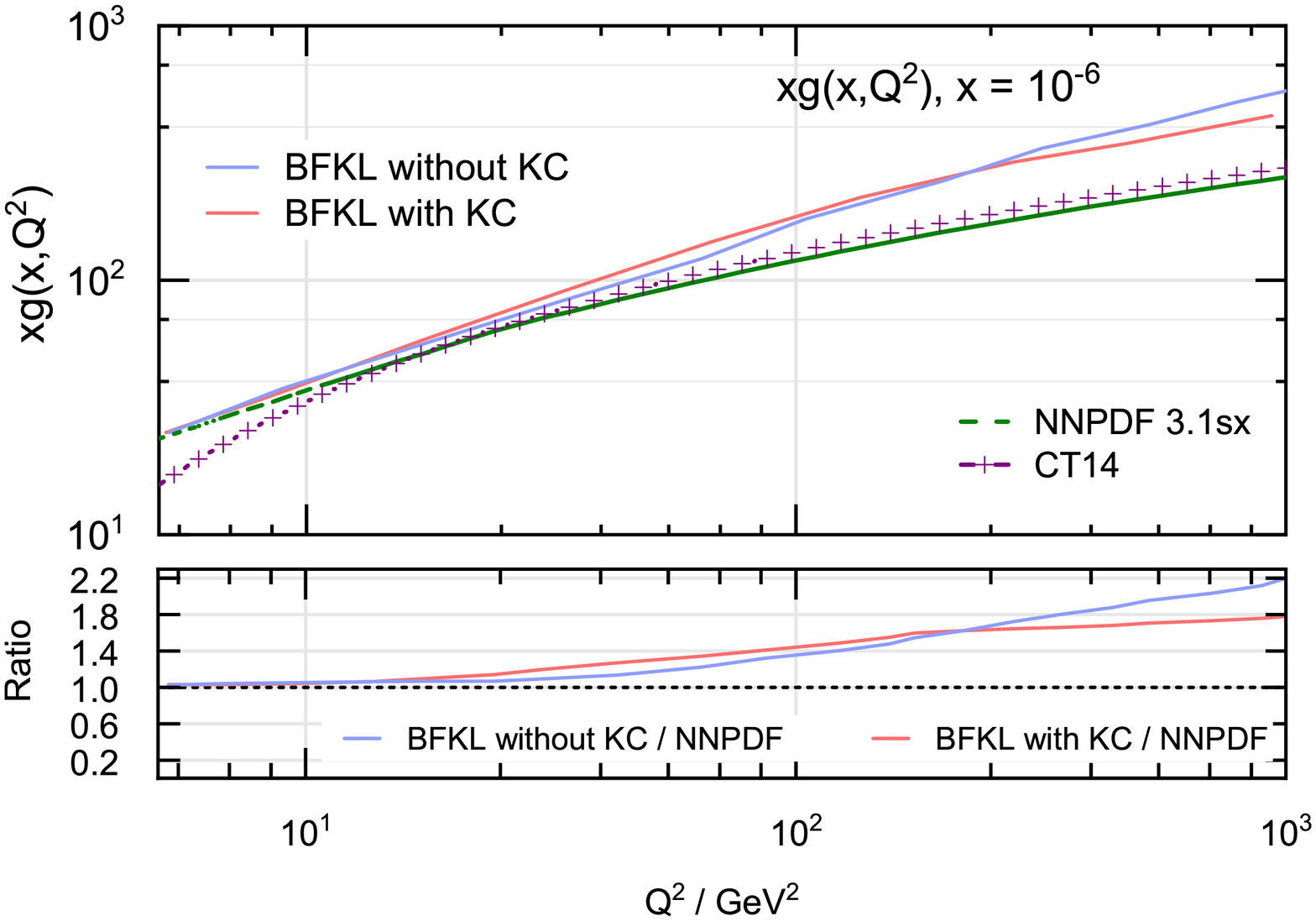}
		
		\caption{\label{f6}$Q^2$  evolution of collinear gluon distribution $xg(x,Q^2)$. Our extracted collinear gluon density from KC-BFKL evolution is contrasted with that of global data fits NNPDF3.1sx and CT14.}
	\end{figure*}

 The unintegrated gluon distribution is related to the conventional gluon density or the collinear gluon density $xg(x,Q^2)$ with the standard relation,
	\begin{equation}
	\label{f2g}
	xg(x,Q^2)=\int_{0}^{Q^2}\frac{\text{d}k_T^2}{k_T^2}f(x,k_T^2).
	\end{equation}
	Using \eqref{f2g} we have extracted collinear gluon distribution from our solution of KC-BFKL equation which is sketched in Fig.~\ref{f5} and Fig.~\ref{f6}. Our predicted collinear gluon density is compared with that of LHAPDF global parameterization groups NNPDF 3.1sx \cite{36} and CT 14 \cite{53}. Both of the LHAPDF datasets  include HERA as well as recent LHC data with high precision PDF sensitive measurements.
	
	\par Our prediction of $x$ evolution of collinear gluon distribution $xg(x,Q^2)$ is obtained for two  $Q^2$ values viz. $\text{GeV}^2$ and $100\text{GeV}^2$ while that of $Q^2$ evolution is obtained for  two $x$ values viz. $x=10^{-3}$ and $10^{-6}$. Our theory is roughly in agreement with data fits for the entire kinematic range of $x$ and $Q^2$. However at very small $x$ regimes our theory predicts slightly faster growth than that of the data fits. This is  expected as BFKL equation account for only gluon splitting and gluon fusion corrections are completely overlooked. To entertain the corrections from gluon fusion at very small $x$ regime, one has to consider nonlinear evolution equations, for instance, Balitsky-Kovchegov equation \cite{10,11} which is beyond the scope of this literature.
	
	\par
\begin{figure*}[t]
		\centering 
		\includegraphics[trim=25 5 21.5 0,width=.46\textwidth,clip]{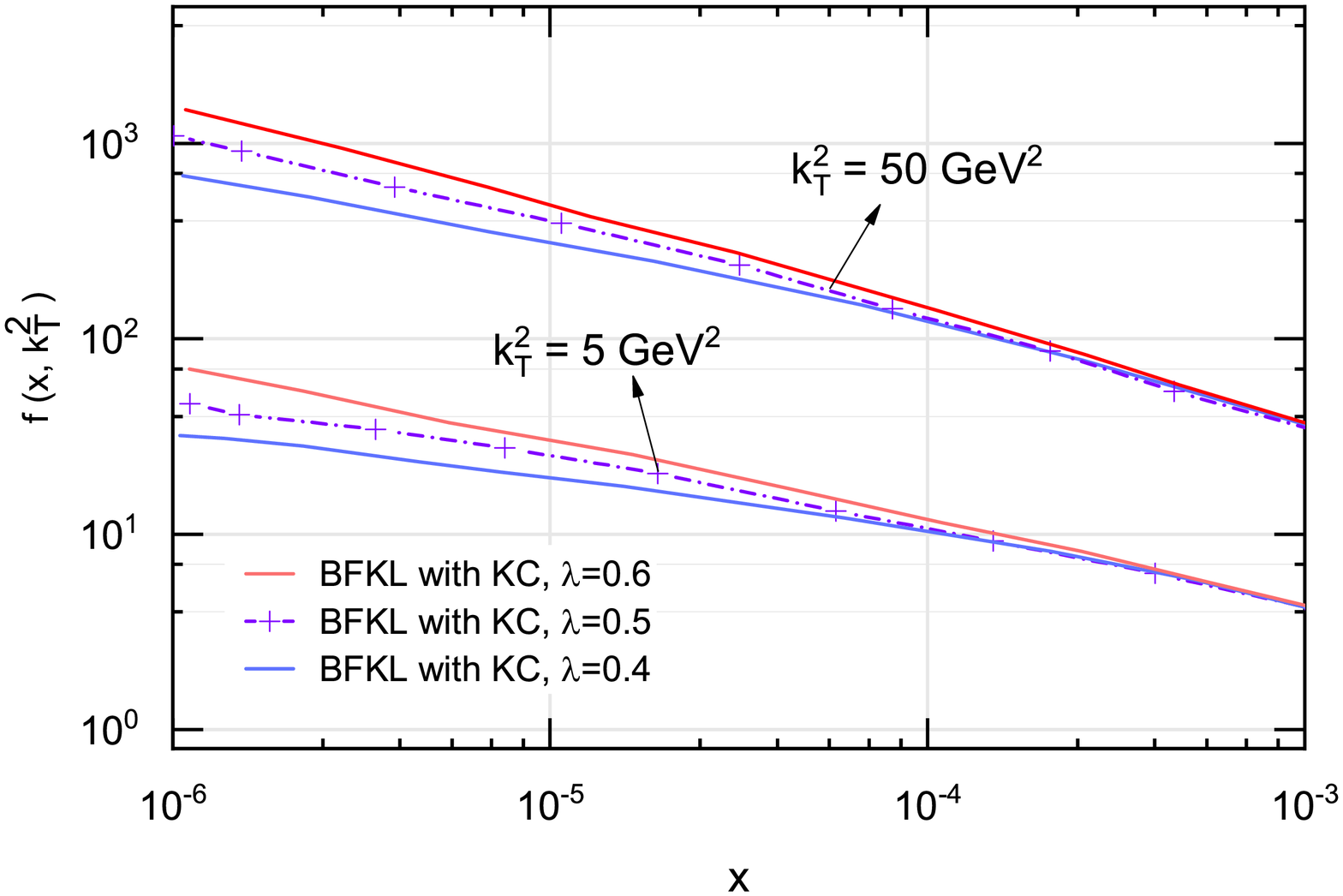}\hspace{5mm}
		\includegraphics[trim=25 5 21.5 0,width=.46\textwidth,clip]{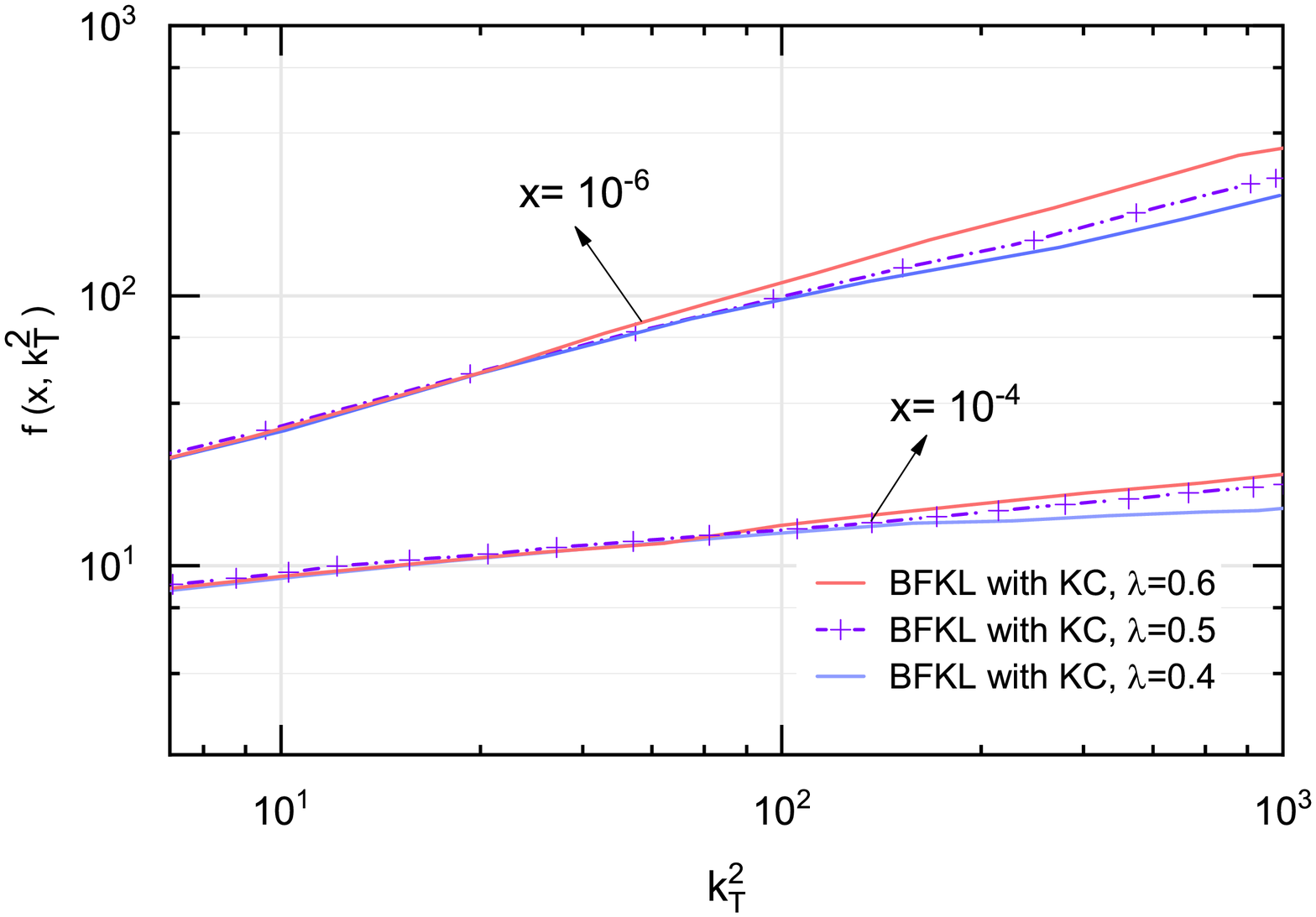}
		
		\caption{\label{f7}BFKL intercept $\lambda$ sensitivity of ugd's $f(x,k_t^2)$ in $x$ evolution (left) and $k_t^2$ evolution (right). }
	\end{figure*}

 We have also performed an analysis (Fig.~\ref{f7}(a-b)) to check the sensitivity of the BFKL intercept $\lambda$ on our solution of KC-BFKL equation. We have sketched both the $x$ and $k_t^2$ evolution for three canonical choices of $\lambda$ viz. $\lambda=0.4$, 0.5 and 0.6 corresponding to three $\alpha_s$ values 0.15, 0.19 and 0.23. Our solution seems to be very sensitive towards a small change in $\lambda$. An apparent 50\% change in $\lambda$ (0.4 to 0.6) suggests atleast around 70-80\% magnitude rise in the gluon distribution $f(x,k_T^2)$ for $x$ evolution. However, sensitiveness of $\lambda$ is comparatively weak in case of $k_t^2$ evolution as it is seen in Fig.~\eqref{f7}. For 50\% change in $\lambda$ (0.4 to 0.6) forecasts approximately around 40-50\% rise in $f(x,k_T^2)$ for $k_t^2$ evolution.

	\section{Conclusion}\label{con}
	In conclusion, we have studied the small $x$ behavior of the gluon distribution particularly in the regime $10^{-6}\leq x\leq 10^{-2}$ and $5\text{ GeV}^2\leq k_t^2\leq 10^{3}\text{ GeV}^2$. In this small $x$ regime, it is necessary to resum the leading logarithmic $\ln(1/x)$ contribution and thereby importance of unintegrated gluon distribution $f(x,k_t^2)$ comes into play. The resummation of leading logarithmic $\ln(1/x)$ is accomplished by the linear BFKL evolution equation. As a consequence of BFKL multi-Regge kinematics, some higher order effects such as kinematic constraint $\theta(k^2/z-k^{'2})$ becomes important, which actually ensures the validity of BFKL equation at small $x$. The major aim of this literature is to explore the effect of kinematic constraint on the small $x$ gluon evolution. Accordingly, in Sect.~\ref{theory} we have implemented the kinematic constraint on the original BFKL equation and able to obtain an integro-differential form of KC supplemented BFKL equation. Using Regge factorization and Taylor's expansion series along with the idea of BFKL multi-Regge kinematics, we have able to express the KC-BFKL equation into an analytically solvable form. Then we have introduced the method of characteristics to solve the PDE and got an approximate analytical solution for both $x$ and $k_t^2$ evolution. In Sect.~\ref{result}, we have studied the $x$ and $k_t^2$ dependence of the ugd's $f(x,k_t^2)$ for the kinematic domain $10^{-6}\leq x\leq 10^{-2}$ and $5\text{ GeV}^2\leq k_t^2\leq 10^{3}\text{ GeV}^2$. Our KC-BFKL 
	evolution is contrasted with the original BFKL equation and a comparatively slower growth of gluon density towards very small $x$ is observed. We have also extracted collinear gluon density $xg(x,Q^2)$ from the ugd's $f(x,k_t^2)$ and drawn a comparison between our theory and global data fits  viz. NNPDF3.1sx and CT14. A rough agreement between our theory and data is obtained for the full domain of $x$ and $k_t^2$ under study. Finally, we have sketched the sensitiveness of the BFKL intercept $\lambda$ in gluon evolution taking three canonical choices of $\lambda$ viz. 0.4, 0.5 and 0.6. A high and moderate sensitivity of $f(x,k_t^2)$ towards the parameter $\lambda$ is seen for $x$ and $k_t^2$ evolution respectively. From these phenomenological studies throughout this literature, we have come to a conclusion that the kinematic constraint effect has a significant impact on small $x$ gluon evolution and should be implemented in any realistic analysis of small $x$ physics.
	\bibliography{111.bib}
	
\end{document}